# Stoner instabilities and Ising excitonic states in twisted transition metal dichalcogenides

Augusto Ghiotto[1†], LingNan Wei[2†], Larry Song[1†], Jiawei Zang[1], Aya Batoul Tazi[1], Daniel Ostrom[1], Kenji Watanabe[3], Takashi Taniguchi[3], James C. Hone[4], Daniel A. Rhodes[5], Andrew J. Millis[1,6], Cory R. Dean[1*], Lei Wang[2*], and Abhay N. Pasupathy[1,7*]

[1]Department of Physics, Columbia University; New York, USA.

[2]National Laboratory of Solid-State Microstructures, School of Physics, Nanjing University; Nanjing, China.

[3]National Institute for Materials Science; Tsukuba, Japan.

[4]Department of Mechanical Engineering, Columbia University; New York, USA.

[5]Department of Materials Science and Engineering, University of Wisconsin; Madison, USA.

[6]Center for Computational Quantum Physics, Flatiron Institute; New York, USA.

[7]Condensed Matter Physics and Materials Science Division, Brookhaven National Laboratory; Upton, USA.

*Corresponding authors. Email: cd2478@columbia.edu, leiwang@nju.edu.cn, apn2108@columbia.edu

†These authors contributed equally to this work.

*Moiré transition metal dichalcogenide (TMD) systems provide a tunable platform for studying electron-correlation driven quantum phases (1). Such phases have so far been found at rational fillings of the moiré superlattice, and it is believed that lattice commensurability plays a key role in their stability. In this work, we show via magnetotransport measurements on twisted WSe$_2$ that new correlated electronic phases can exist away from commensurability. The first phase is an antiferromagnetic metal that is driven by proximity to the van Hove singularity. The second is a re-entrant magnetic field-driven insulator. This insulator is formed from a small and equal density of electrons and holes with opposite spin projections – an Ising excitonic insulator.*

Moiré superlattice patterns imposed on transition metal dichalcogenide (TMD) semiconductors such as $WSe_2$ create potential wells for electrons and holes in each moiré unit cell. Motion of electrons in the geometry of the lattice defined by the wells creates a bandstructure of extended electronic states that is tunable by twist angle and applied electric fields and hosts novel phases of matter (*2, 3*). Recently, moiré patterned TMDs have been found to display a multitude of insulating phases at rational fillings of the moiré lattice (*1*). These phases may be understood as arising from a combination of the potential energy that localizes individual electrons within the moiré unit cell and the strong Coulomb repulsion between electrons with the specifics of the bandstructure being unimportant. In contrast, other familiar interaction-driven phases in metals or doped semiconductors feature an interplay between the band structure describing extended states and the Coulomb repulsion between electrons. A classic example is the ferromagnetic Stoner instability (*4*).

Here, we report magnetotransport measurements on hole-doped twisted bilayer $WSe_2$ (t$WSe_2$ henceforth) at moderate twist angles that indicate an array of tunable metal-insulator and metal-metal phase transitions driven by an interplay of correlation and band structure effects. Going beyond previous work showing evidence of Mott insulating phases at half- filling of the first moiré band (*1, 5, 6*), we find effects beyond the Mott paradigm, often occurring away from half filling and driven by proximity to the van Hove singularity (vHs) or, at high spin polarization, by an electron-hole interaction.

Our measurements are performed on a t$WSe_2$ sample with a 4.2° twist and remarkably low disorder. Samples with twist angles larger than about 5° are difficult to gate due to limitations of gate dielectric strength while samples with twist angles much less than 3° suffer from stronger disorder relative to the bandwidth. We fabricate our samples in a dual-gated structure to allow for independent control of carrier concentration and displacement (perpendicular electric) field (*5, 6*). The highest valence band of t$WSe_2$ may be described as a single orbital tight binding model defined on the sites of the triangular moiré superlattice with a bandwidth comparable to interaction scales and with relatively weak disorder. Importantly, both the electron density and the band structure of a single sample can be sensitively tuned over wide ranges by the gate voltages. Further, the large moiré unit cell size means that at experimentally accessible fields it is possible to insert appreciable fractions of a flux constant per unit cell while the large Landé g-factor intrinsic to $WSe_2$ (*7, 8*) provides a unique opportunity to study physics in the regime of high spin polarization (*9*).

We measure the longitudinal and Hall resistivity as a function of magnetic field and bottom gate voltage, for several fixed values of top gate voltage. The difference between top and bottom gate voltage determines the "displacement field" D which controls the electronic band structure (*5, 6*) while the average controls the carrier density, which is independently determined as discussed in the methods section (see SI).

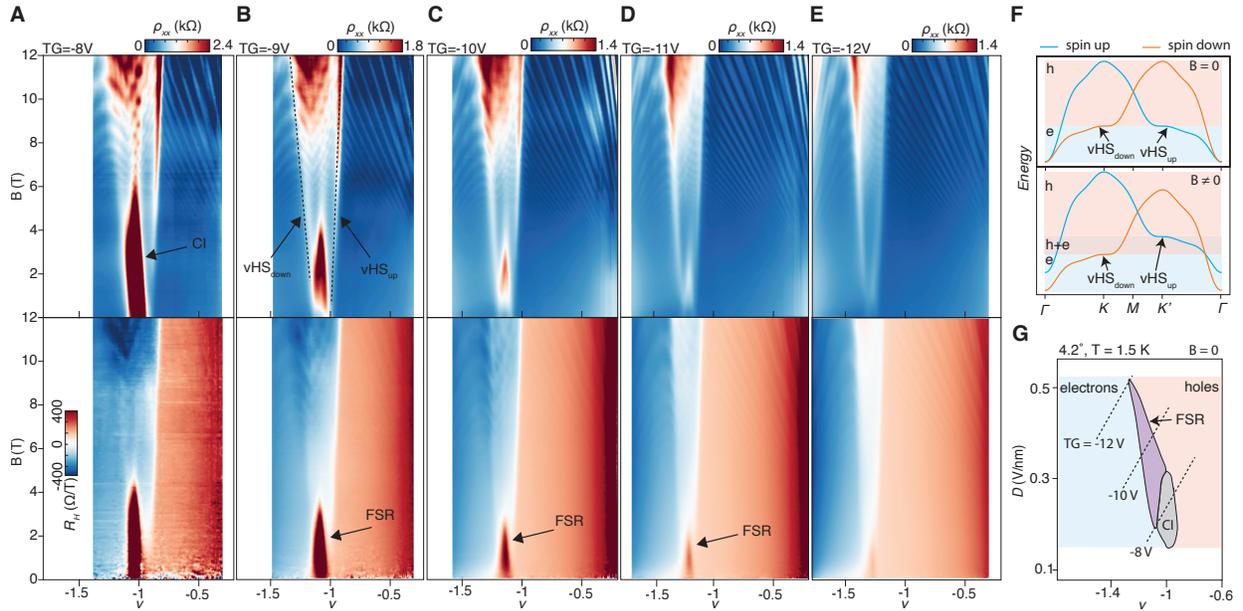

**Fig. 1. Fermi surface reconstruction and magnetic-filed tunable van Hove Singularities**
**(A-E)** Longitudinal resistivity $\rho_{xx}$ (top panels) and Hall coefficient $R_H$ (bottom) as a function of magnetic field B and carrier density ν (determined as described in the supplementary material) for different top gate voltages. The $\rho_{xx}$ plots reveal that for certain top gate values and at low magnetic fields, narrow regimes of high resistivity correlated insulator (CI) phase separate wide regions of low resistivity metallic behavior and low resistivity metallic regimes, while for other top gate values no CI behavior is observed. The white slanted lines are very sharply defined regions of moderately high resistance behavior across which the Hall resistance ($R_H$, bottom panels) changes abruptly, justifying their interpretation as locations where the spin up and spin down Fermi surfaces pass through van Hove points in the band structure (panel F). In the vicinity of the point to which the van Hove lines extrapolate at B->0 a region of very high Hall resistance is observed, even for top gate values at which no CI phase is found. The change in Hall effect is interpreted as arising from a Fermi surface reconstruction (FSR). **(F)** Sketch of the valence bands in tWSe$_2$ at a nonzero displacement field. At zero magnetic field, the bands are degenerate under simultaneous exchange of spin and valley (upper panel). When a magnetic field B is applied, the symmetry is broken via Zeeman coupling to B (rightmost). **(G)** Zero magnetic field phase diagram inferred from the results of panels A-E. Dashed lines show the trajectory in the plane of carrier density and displacement field obtained by varying the bottom gate voltage at the fixed top gate values indicated. The pink and blue colors indicate regions of hole and electron-like metallic transport as determined from the Hall resistance; these regimes are separated by regimes of correlated insulator (smaller TG) or Fermi surface reconstructed metallic behavior (larger TG).

Figures 1A-E display color maps of the longitudinal (top) and Hall (bottom) resistances for different magnetic fields applied perpendicular to the sample plane as the back gate voltage (parametrized by the electron density ν) is swept while the top gate is maintained at a constant

value, corresponding to a simultaneous change in displacement field and carrier concentration. All of the Hall data in this work has been antisymmetrized and the longitudinal resistance data has been symmetrized in magnetic field. For a top gate (TG) voltage set at -8 V (fig. 1A), a correlated insulator is present at half-filling (*5, 6*). We identify the phase as a correlated insulator from the temperature dependence of the resistivity (fig. S2). For TG = -9 to -12 V (figs. 1B-E at zero magnetic field the sample remains metallic at all points on these plots as seen from the magnitude of the resistivity and from the temperature dependence shown in fig. S2.

The field dependence of the transport gives us insight into the bandstructure of the system. At high field, Landau levels are observed both in the longitudinal and Hall responses. These levels emanate both from the top of the valence band (hole-like) and from full-filling of the first moiré band (electron-like). Also evident on all these plots at high field are two additional lines on the longitudinal response (dashed lines in fig. 1B). These lines separate two different regimes of Landau fan structure and indicate local maxima in the resistance. Across each of these lines, sharp steps are seen in the Hall response. The change in the Hall response, especially the sign change observed across the line at $\nu > -1$ shows that these two lines correspond to the van Hove singularities of the majority and minority spin channels, which are Zeeman split at nonzero field. We see that the magnitude of the Zeeman splitting can be large even at modest fields of 12 T, corresponding to a filling difference of 0.3-0.4 holes per moiré unit cell. Further, for top gate values of $-9, -10$ and (weakly) $-11$V and at modest fields we see a small region of very large magnitude hole-like Hall response centered at a top gate-dependence carrier concentration and not necessarily tied to regions of insulating behavior. As discussed in more detail below we associate this behavior with a Fermi surface reconstruction which we attribute to itinerant antiferromagnetism. Fig. 1F sketches the bandstructure of tWSe$_2$ and the Zeeman effect of an external magnetic field in splitting the bands. Fig. 1G shows the phase diagram at zero magnetic field in the plane of doping ($\nu$) and displacement field (D) with the cuts obtained by varying carrier concentration at fixed top gate voltage shown as dashed lines. The zero-field correlated insulator (CI) exists for a narrow range of D from $\sim 0.15$ to $\sim 0.3$ V/nm at $\nu = 1$ and the Fermi surface reconstruction (FSR) for a wider range of D and $\nu$ generically different from 1.

**Correlated Insulator and Fermi surface reconstruction**

We now examine in more detail the response at modest fields (up to 5 T) in figs. 1A-E. First, we consider fig. 1A where the correlated insulator is present at zero field. Upon applying a perpendicular magnetic field, the insulator transitions into a metal at ~5 T. The corresponding Hall resistance $R_H = \rho_{xy}/B$ (fig. 1A, bottom) is also high in the region of the CI at low fields. We believe that in an ideal sample the zero-field state observed at TG= $-8$V would be an ideal non-topological insulator at hole density $\nu = -1$ and that the nonzero values seen are a consequence of thermal excitations and the small gaps in the experimental sample.

For all of the other top gate voltages studied, the B=0 longitudinal resistivity remains metallic, as shown by the temperature dependences reported in Fig. S2, albeit with a weak maximum at a top-gate dependence hole concentration greater than 1. However, the extrapolation to zero field of the Hall resistance exhibits a region of large hole-like and field-dependent Hall resistance as the top gate is increased further towards the maximum value of -12 V (figs. 1C-E), the low-field anomalous behavior becomes progressively less prominent. As discussed in more detail below we use this feature of the Hall resistance to define the region of Fermi surface reconstruction (FSR) region of the B=0 phase diagram shown in fig. 1G. We observe that the CI is restricted to near $\nu = -1$, while the FSR is measured at a top-gate dependent range of incommensurate fillings.

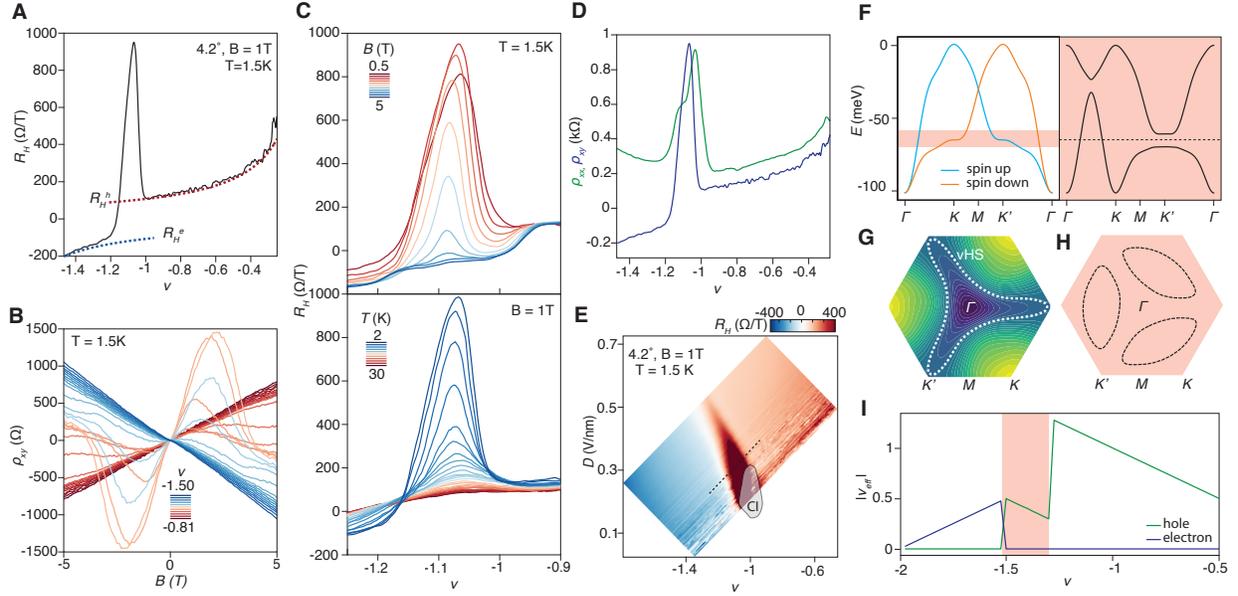

**Fig. 2. Incommensurate spin density wave in the metallic state**

**(A)** Hall coefficient $R_H$ (solid black line) taken at B=1 T and temperature 1.5 K for TG = -9 V. Doping-based expected values of $R_H$ for hole-like (red dashed) and electron-like (blue dashed) behavior are also plotted. The data agree with the expected values except in the vicinity of $\nu = -1$, where there is a near tenfold increase in $R_H$. **(B)** $\rho_{xy}$ versus magnetic field for the doping range shown in a. Dopings well away from $\nu=-1$ (darker red and blue lines) exhibit an essentially linear field dependence consistent with the band structure carrier density, over the whole field range. Dopings close to $\nu=-1$ (lighter blue and red lines), a large amplitude linear $\rho_{xy}$ behavior at low fields reverting at high fields to the expected band theory behavior. **(C)** $R_H$ versus doping as a function of magnetic field (top) and temperature (bottom). At fields greater than 4 T and temperatures above 10 K, the peak in $R_H$ vanishes. **(D)** $\rho_{xx}$ taken at 0 T and $R_H$ taken at 1 T versus doping for TG = -9 V. **(E)** Displacement field D and doping dependence of $R_H$ taken at 1 T. The $R_H$ peak decreases at higher displacement field and skews towards higher dopings, overlaid for reference with the CI region of the phase diagram. **(F)** Single particle electronic bands of tWSe$_2$ at a non-zero displacement field (left) and reconstructed bands driven by spin density waves near the vHs (right) displaying small hole pockets. The shaded red region indicates the energy range where a Fermi surface reconstruction due to a weak antiferromagnetic instability significantly affects the band structure. **(G)** Single particle spin up Brillouin zone of tWSe$_2$. White dashed line indicates the vHs. **(H)** Reconstructed Fermi pockets near the vHs. **(I)** Theoretically expected density of holes and electrons as function of filling $\nu$.

We examine in fig. 2 the density, field and temperature dependence of the Hall coefficient following previous analyses of Fermi surface reconstructions in the cuprates (*10*) and heavy fermions (*11*). Fig. 2A shows a density line cut from fig. 1B (top gate voltage -9 V) at magnetic field B = 1T and low temperature T = 1.5K, compared to the Hall resistance obtained assuming a single carrier type using the classical formula $R_H = 1/ne$, with n the appropriate carrier density based on the gate voltage values. The classical Hall resistance is expected to be monotonic and smoothly interpolate through zero at the van Hove singularity. The measured data agrees well with the classical values except for ν between -1.2 and -1, where a strong hole-like peak in $R_H$ is measured.

The field dependence of the Hall resistance at different carrier densities (fig. 2B) gives us further information. At doping levels where the Hall resistance in fig. 2A matches the non-interacting band theory expectation, the Hall resistance is linear in applied field, as expected from single electron or hole pockets. In the region where the $R_H$ peak is measured ($-1.2 < \nu < -1$), field dependence exhibits pronounced nonlinearities with applied field. A linear field dependence is first seen at low fields with a large positive slope, indicating the presence of small hole pockets since $R_H = 1/ne$. At fields above ~2 T, $\rho_{xy}$ passes through a maximum, decreases, and changes sign and begins to converge to electron-like behavior. To better visualize the effect of magnetic field on the $R_H$ peak, fig. 2C displays $R_H$ versus doping for different B fields (top) and different temperatures (bottom). We notice that the peak gradually vanishes with increasing perpendicular magnetic field and temperature, disappearing above B $\gtrsim$ 4 T or T $\gtrsim$ 10K.

The density and field dependence described above are consistent with a Fermi surface reconstruction that occurs between $-1.2 < \nu < -1$ at low B and low T. The sensitivity to magnetic field and the disappearance of the nonlinearity in the Hall effect above ~3 T indicate that the FSR is of antiferromagnetic nature. To get further evidence that the behavior described above is a true phase transition, we measure the temperature dependence of the low-field Hall resistance, as shown in fig. 2C (bottom). This data shows that the additional peak in the Hall resistance that signals the FSR weakens with increasing temperature, and at temperatures above ~15 K, the Hall effect across the entire doping range is well-described by non-interacting electron theory.

Along with the signatures of the FSR in the Hall response, the longitudinal resistance also shows signatures of the FSR in its temperature dependence (see fig. S2). Fig. 2D shows that, similarly to materials that exhibit charge and spin density waves (*12, 13*), a kink in $\rho_{xx}$ is seen at dopings similar to when a peak in $\rho_{xy}$ emerges.

The region of the FSR is summarized in fig. 2E, which shows the Hall response as a function of both density and displacement field. Temperature dependence of the longitudinal resistance determines the region of the correlated insulator, which is shown by the gray shaded region on the figure. We see that the FSR is strong when the bands are less dispersive at smaller displacement field. At higher displacement fields, the region of the FSR shrinks and almost disappears at the highest gate voltages we can achieve.

Stoner's criterion states that a high fermi surface density of states and moderate interactions can lead to magnetic ordering. A van Hove singularity, which generically occurs in two dimensional materials, leads to a region of high density of states where magnetic orders are predicted to manifest (*14–19*). In tWSe$_2$, high displacement fields are theoretically predicted (*5*) to make the bands more dispersive as the vHS moves away from half-filling towards full-filling; this and the larger size of the reconstructed pockets means the FSR effect weakens as the doping is moved away from ν=-1. Shown in fig. 2F (left) are the moiré valence bands of tWSe$_2$ at a nonzero displacement field, obtained from DFT calculations. When the Fermi level is near the vHS (shown in red shade), antiferromagnetic ordering reconstructs the bands as shown in fig. 2F (right), obtained from mean field calculations assuming a commensurate 120° spiral order. Mean field theory becomes more complicated for incommensurate orders and will be discussed elsewhere. When the Fermi surface is reconstructed, electron and hole pockets can coexist, depending on displacement field and the carrier concentration. Fig. 2G shows the single particle spin-up Brillouin zone and the vHS is denoted in a white dashed line. Fig. 2H shows the reconstructed Brillouin zone when the Fermi level is near the vHS (as shown in fig. 2F (right)) and small hole pockets are left after folding of the bands. Fig. 2I indicates the effective density for the moiré valence band. When the Fermi is reconstructed (red shaded region), there is a sudden drop in the hole density and, and thus a peak in $R_H = 1/ne$ is expected, which agrees with our measurements.

**Ising excitonic states at high magnetic fields**

We have so far discussed correlation-induced instabilities at weak magnetic fields, where the physics was of a van Hove singularity-induced translational symmetry breaking leading to a Fermi surface reconstruction. In what follows, we consider a different instability, occurring at very high fields and near half filling, which we interpret as an excitonic insulator occurring in a system with a small and approximately balanced number of electrons and holes. Previously, several excitonic insulator candidates have been found experimentally that can be divided into interband systems such as $TiSe_2$ (*20*), $WTe_2$ (*21, 22*) and $Ta_2NiSe_5$ (*23–25*); and interlayer systems such as GaAs bilayers, (*26*), bilayer graphene/hBN/bilayer graphene (*27, 28*), $WSe_2/hBN/MoSe_2$ (*29*) and bilayer $WSe_2$ (*30*). More broadly, other excitonic phases beyond insulators can exist in systems where electrons and holes coexist. They include excitonic metals, where the large density of excitons screens their interactions and prevents condensation (*27, 28*) and imbalanced excitonic metals with excess electrons or holes that give rise to topological states at high magnetic fields such as those recently observed in InAs/GaSb quantum wells (*31, 32*).

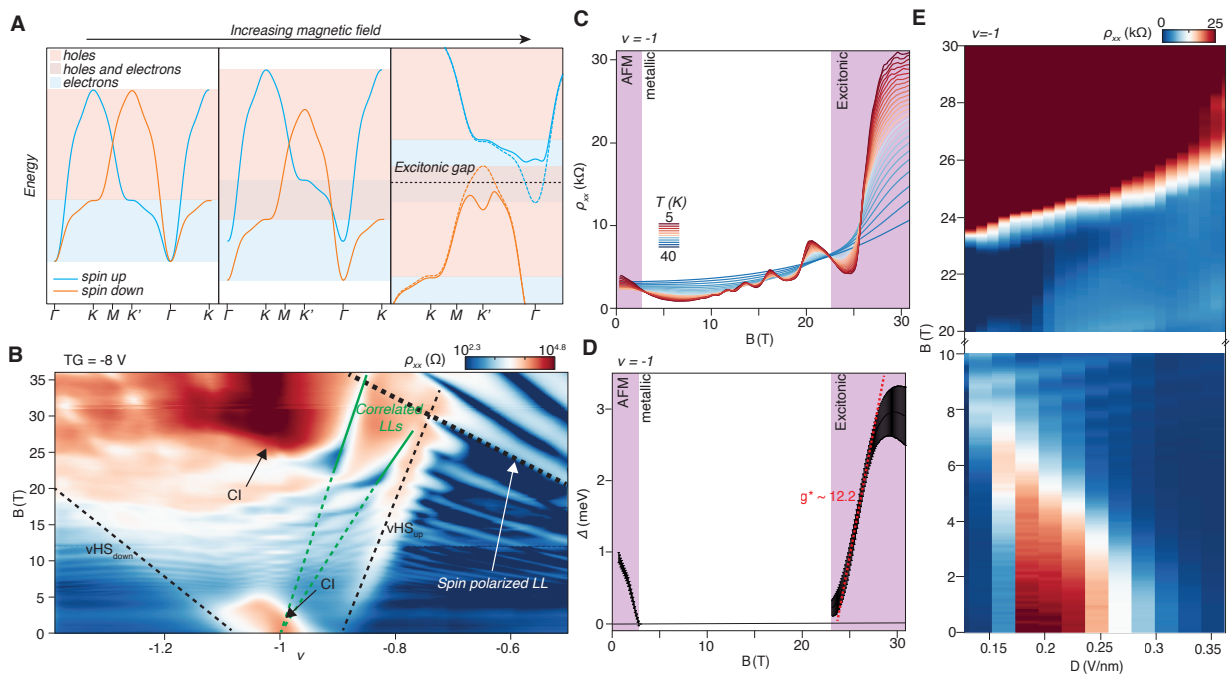

**Fig. 3. High magnetic field re-entrant correlated insulator**
**(A)** At sufficiently high magnetic fields, an overlap of the electron pocket of the spin-valley up subband with the hole pocket of the spin-valley down subband can be realized. In the regime of moderate interactions, an excitonic insulator can be realized. **(B)** Longitudinal resistivity $\rho_{xx}$ versus filling and magnetic field at TG = -8V. CIs are pointed with an arrow as well as the spin-valley up and down vHSs as dashed black lines. High field correlated LLs emanating from $v =$

−1 are highlighted in green and the dashed black line indicates the field at which the occupied Landau levels become fully spin-polarized LL. **(C)** Longitudinal resistivity $\rho_{xx}$ versus magnetic field at ν = −1. Increasing magnetic field above 5T destroys the antiferromagnetic insulator and eventually leads to a re-entrant insulator above 24 T, which is interpreted as an excitonic insulator. Temperature scale is set in even spaces of 1/T. **(D)** Field dependence of the low field and high field insulators. An effective g-factor of about 12.2 is extracted for the excitonic insulator from the magnetic field dependence of the Arrenhius gap in the resistivity **(E)** Magnetic field CIs phase diagram as a function of displacement field.

We achieve the regime of simultaneously small electron and hole densities in tWSe$_2$ by applying a large perpendicular magnetic field. The Zeeman effect results in an imbalance between spin up and down carriers. Due to the large Landé g-factor, ranging between 20 to 30 for monolayer and Bernal bilayer (*7, 8*), we can significantly tune the imbalance between spin up and spin down carriers as shown in fig. 3A. In fact, for the parameters of our sample, a field larger than 50 T is sufficient to fully spin polarize the band at a total density of ν = −1 (*17*). At fields that are close to but smaller than this number, the spin-down band has a small population of holes while the spin-up band has a small population of electrons. At a total density of ν = −1 the density of holes and electrons is enforced to be identical, and both types of carriers live in Landau levels at high applied magnetic fields. These are ideal conditions for the formation of an excitonic insulator (*17, 33*).

Shown in fig. 3B is the longitudinal resistance at T = 1.5 K as a function of density and magnetic field at a top gate of -8 V. We first focus on the behavior at a total density of ν = −1. The behavior of the resistance as a function of magnetic field is shown for several temperatures at this density in fig. 3C. At zero field, the system is a correlated insulator with a small gap of 1 meV as we have described before (*5, 6*). This insulator is destabilized by a small magnetic field (∼5 T), leading to a metallic temperature dependence at moderate fields. Further increasing the magnetic field leads to another metal-insulator transition above ∼24 T, as shown in fig. 3A. Arrhenius fits to resistivity ($\rho \sim e^{\Delta/2k_BT}$) allow us to extract a field dependent gap for both the low field and high field insulators (fig. 3D). The gap at high field grows approximately linearly with magnetic field up to about ∼28 T, where it stops increasing. A linear fit of the high-field gap $\Delta(B) = g*\mu_B(B − B_c)$ yields $g* = 12.2$.

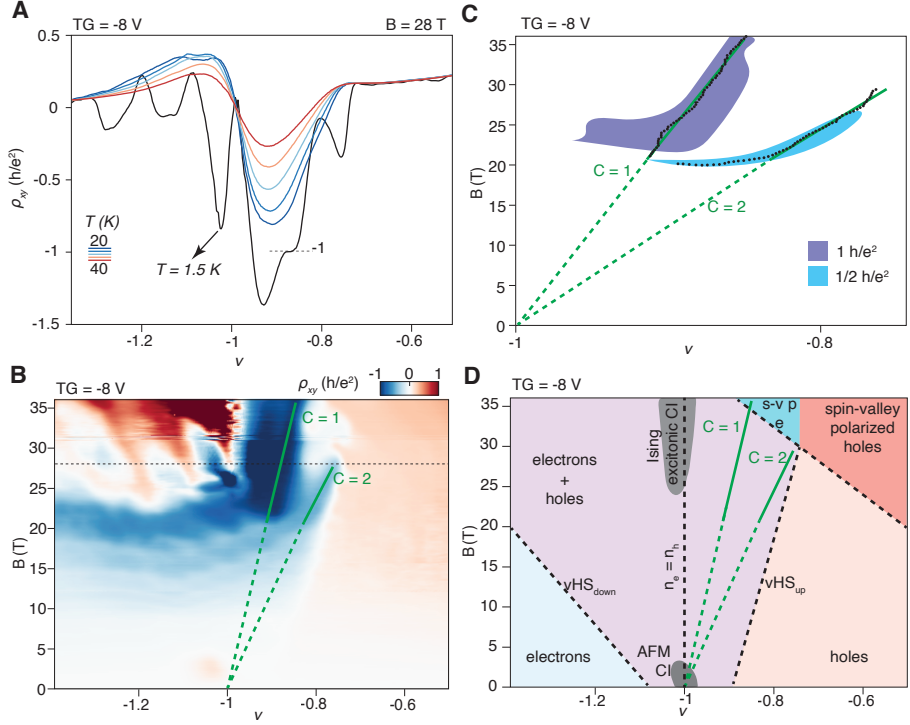

**Fig. 4. Topological excitonic states in the electron-imbalanced regime**
**(A)** Density dependence of $\rho_{xy}$ at fixed B field of 28 T and temperatures indicated where the insulator is present. The black curve was taken at 1.5 K, where a $h/e^2$ quantum Hall plateau develops. In contrast to the region around $\nu = -0.6$, $\rho_{xy}$ exhibits a large temperature dependence in the region where electrons and holes coexist. **(B)** Transverse resistivity $\rho_{xy}$ as a function of doping and magnetic field for TG = -8V taken at 1.5 K. **(C)** Electron-like Landau levels emanating from $\nu = -1$. Dotted lines are the local minima of $\rho_{xx}$ and the shaded regions are the measured values of $\rho_{xy} \pm 0.5\%$. Green lines are the expected slopes from Streda's formula for Chern numbers 1 and 2. **(D)** Low temperature phase diagram in doping and field for TG = -8 V.

One trivial possibility for insulating behavior at high field is if the applied field is large enough to completely spin polarize the band, leading to a gap at a filling of $\nu = -1$. We have three sets of experimental facts that show that this is not the case at fields up to 36 T. These are the magnetic field dependence of the high field insulating gap; the behavior of the Landau level spectrum at low density; and the behavior of the Hall resistance as a function of density. We discuss each of these in turn.

The magnetic field dependence of the high field gap is inconsistent with that expected from a fully spin polarized bands. Our experimental data shows a linear dependence of the gap with magnetic field when it first opens, but with a slope that would correspond to an effective Landé g-factor g* that is much lower than what would be expected from the Zeeman effect. The saturation of the gap for fields larger than 28 T is also inconsistent with complete spin polarization of the bands.

The Landau level structure at low density gives us information on the spin polarization in the system. In particular, near the band edge, full spin polarization exists in the Landau levels at high magnetic field. As the density is increased and the minority spin channel Landau levels pass through the chemical potential, a distinctive even-odd structure emerges in the Landau level spectrum (*8*). We can use this to determine the region in doping and magnetic field where the bands are fully spin polarized (black dotted line, fig. 3D). An extrapolation of this line to $\nu = -1$ indicates that the field for full spin polarization is approximately 40 T.

We have so far discussed the behavior at $\nu = -1$ where the number of electrons and holes is identical. We now consider densities away from this point. In general, when a single carrier type is present, the sign of the Hall effect is set by the carrier type, and no temperature dependence is observed. Shown in fig. 4A is $\rho_{xy}$ as a function of filling at TG = -8V and field 28 T for several values of temperature. At densities $\nu > -0.6$ and $\nu < -1.3$, $\rho_{xy}$ is temperature independent, as is expected classically from a single carrier type. However, between these two values, $\rho_{xy}$ has pronounced temperature dependence, consistent with the presence of both electron and hole carriers. For $-1 > \nu > -1.3$, the $\rho_{xy}$ at 1.5 K changes sign multiple times as a function of filling, indicating strong competition between electrons and holes. Both of these facts illustrate clearly that between 24 and 36 T the sample is not fully spin polarized up between $-2 < \nu < -0.9$.

Shown in fig. 4B is $\rho_{xy}$ as a function of density and field at TG = -8 V at low temperature T = 1.5 K. A surprising feature of this data is the presence of a near-perfectly quantized Hall plateau at low temperature with Chern number 1 for $\nu > -1$ (shown in the solid green lines in fig. 4B). A phenomenological picture to explain this observation is that the spin down holes continue to be bound to the spin up electrons in this region, giving no contribution to the Hall effect. The residual spin up electrons then give a quantized Hall effect. Recently, similar topological states have been observed in the electron imbalanced excitonic insulator InAs/GaSb (*31, 32*). In fig. 4C, we map out the areas where a plateau in $\rho_{xy}$ was observed in the electron-imbalanced region of the excitonic metal within 0.5% of the quantization value. There are two prominent plateaus, one at 1 $h/e^2$ (shaded in navy blue) and the other at 1/2 $h/e^2$ (cyan). On the same figure, we juxtapose the local minima of $\rho_{xx}$ (black dots) and the expected slopes of the first two LLs emanating from $\nu = -1$ according to Streda's formula n = CeB/h, where C is their Chern number.

The correlated topological states seen for $\nu > -1$ are not detected for $\nu < -1$. Indeed, $\rho_{xy}$ fails to quantize at a temperature of 1.5 K as seen in fig. 4A. Instead, patches of hole-like and electron-like behavior compete. This is exemplified in the line cut shown in fig. 4B, where, at 1.5 K, $\rho_{xy}$ crosses zero multiple times (see fig. S1). In the excitonic picture, this implies that the exciton binding is destroyed quickly for $\nu < -1$. Such differences can come from the inherent electron-hole asymmetry in the bandstructure on doping away from $\nu = -1$. Finally, we map out the rich magnetic field vs. doping phase diagram at low temperature for TG = -8 V in fig. 4D.

This species of excitonic insulator, which we here call "Ising excitonic insulator", has a unique character as it involves binding of $S_z=+1/2$ electrons and $S_z=-1/2$ holes. Such a situation is difficult to realize in graphene with laboratory magnetic fields due to the g-factor being close to 2. We note that a re-entrant insulator at half-filling has been observed in magic angle graphene near one flux quantum and it is argued that the flatbands renormalizes when the magnetic flux is comparable to the moiré unit cell area (*34*). This is not the case in our system, where one flux quantum corresponds to 238.4 T.

**Conclusion**

Our experimental findings shed important light on the physics of interactions in twisted $WSe_2$ samples. Primary among them is the relationship between the Fermi surface reconstructions discovered here and the insulating states that have previously been discovered at $\nu = -1$. Our current data indicates that the signatures of the Fermi surface reconstructions become stronger as the van Hove singularities approach $\nu = -1$ from below. This is consistent with a picture where both the correlated insulator and the metallic phase immediately adjacent to it for $\nu < -1$ are both magnetically ordered. Interestingly, we do not see clear signatures of the Fermi surface reconstruction for $\nu > -1$. Whether this is a limitation of transport measurements or whether the metal-insulator transition for $\nu > -1$ is of a fundamentally different origin remains an open question. Our measurements at high field show that $tWSe_2$ can be brought to a very different electronic regime with laboratory magnetic fields. The observation of strong excitonic correlations in a single two-dimensional material (rather than in spatially separated layers) opens possibilities for achieving strongly correlated excitonic phases beyond those observed so far in $tWSe_2$, such as excitonic superfluids. However, the fact that these phases occur in a single material makes it more difficult to apply techniques developed for spatially separated layers such as counterflow measurements. The development of new techniques to probe these phases at high field is an interesting experimental challenge.

**Acknowledgments:** The authors thank Chandra Varma, T. Senthil, Patrick A. Lee, Liang Fu, Debanjan Chowdhury, Eun-Ah Kim, Shaffique Adam, Dante M. Kennes, Jennifer Cano, Feng Wang, Philip Kim, Jie Wang and Daniel Parker for fruitful discussions. The authors also thank the staff at NHMFL for the thorough support.

**Funding:**

Programmable Quantum Materials, an Energy Frontier Research Center funded by the U.S. Department of Energy (DOE), Office of Science, Basic Energy Sciences (BES), under award DE-SC0019443. (AG and AJM in part).

Natural Science Foundation of Jiangsu Province Grant No. BK20220066. (LW)

CRD is supported by the Columbia MRSEC on Precision- Assembled Quantum Materials (PAQM) - DMR-2011738. (CRD)

Air Force Office of Scientific Research via grant FA9550-21-1-0378 and the National Science Foundation via grant DMR-2004691. (ANP)

A portion of this work was performed at the National High Magnetic Field Laboratory, which is supported by National Science Foundation Cooperative Agreement No. DMR-2128556* and the State of Florida.

The Flatiron Institute is a division of the Simons Foundation.

**Author contributions:**

Sample fabrication: LNW and LW

Transport measurements: AG, LNW, LS, ABT, DO

Data analysis: AG, LS

WSe$_2$ synthesis: DAR

hBN synthesis: KW, TT

Theoretical modeling: JZ, AJM

Manuscript writing with input from all authors: AG, LNW, LS, LW, ANP, AJM, CRD

Funding acquisition: JCH, AJM, CRD, LW, APN

**Competing interests:** Authors declare that they have no competing interests.

**Data and materials availability:** Raw data is available upon request.

**Materials and Methods**

Sample preparation

We prepared our samples as described in in Wang et al. (*1*). A graphite back gate is picked up with hBN via the dry stamp technique (*2*). Pre-patterned 2/20 nm Cr/Pt contacts were then evaporated on top of the hBN (*3*). Another hBN is used to "tear-and-stack" WSe$_2$, which is then dropped onto the Pt contacts (*4*). Cr/Au is then evaporated as the top gate.

Twist angle, displacement field and doping determination

We determine the geometric capacitance of the top and bottom gates via the Landau level separation in gate voltage and n = NeB/h, where n is the carrier density, N is the filling, e the electron charge, B the magnetic field and h Planck's constant. We then use the projection of the hole-like Landau fan to zero field to determine the intrinsic doping of the sample; and the projection of the electron-like fan to determine the full-filling density. For the sample shown in the paper, $\nu$ = 0.01x($V_{TG}$+$V_{BG}$)+0.09 and D = 0.01x($V_{BG}$ - $V_{TG}$) and full-filling density $n_s$ = -1.161x10$^{13}$ cm$^{-2}$. We find the twist angle using $n_s$ = 2/A, A = $\frac{a^2\sqrt{3}}{4(1-cos\theta)}$ and a = 0.328 nm (*5*).

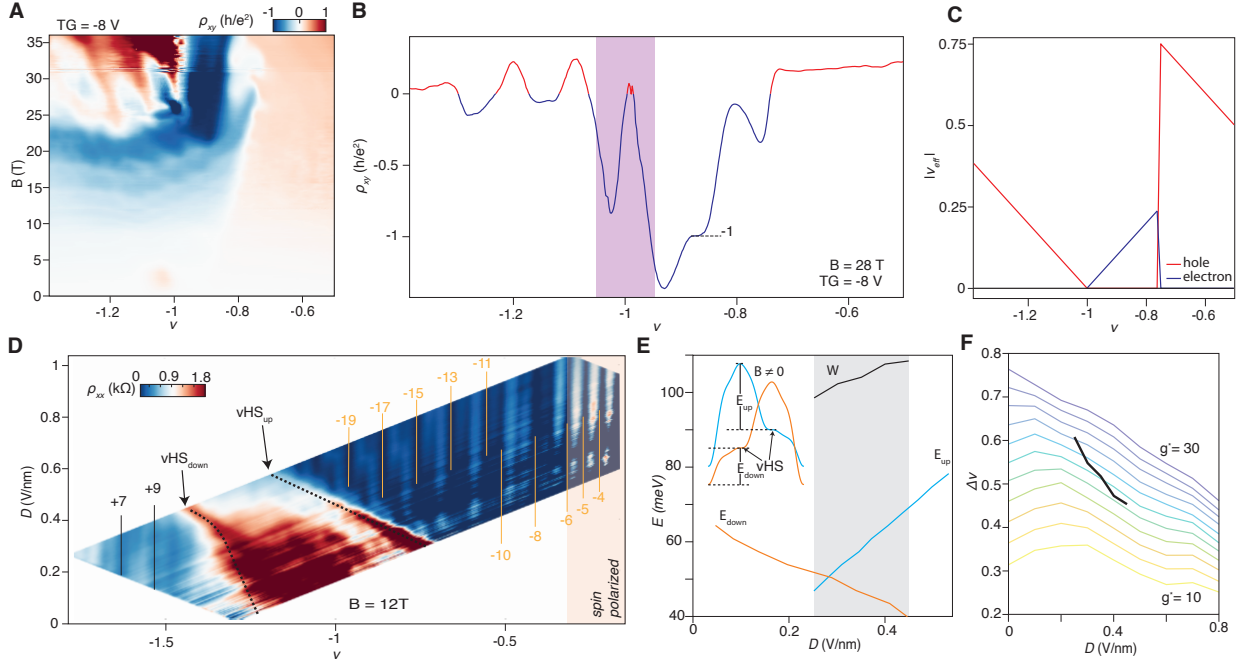

**Fig. S1. Evidence for overlapping bands**
(**A**) Transverse resistivity $\rho_{xy}$ as a function of doping and magnetic field for TG = -8V. (**B**) Line cut of a at 28 T where the insulator is present. Competing hole and electron behaviors emerge below $\nu = -1$. (**C**) Expected populations of holes and electrons from single particles calculations once a trivial gap opens for the range of fillings measured in (A). (**D**) $\rho_{xx}$ versus doping and displacement field at 12 T. Hole-like and electron-like set of LLs are highlighted as well as the positions of the spin-valley up and down vHSs. Hole-like and electron-like LLs are clearly defined up until they cross the vHSs. A well-defined more resistive region between the vHSs where electrons and holes coexist is also present. (**E**) Displacement field dependence of the energy of the up spin $E_{up}$ and spin down $E_{down}$ vHSs (shown in (A)) obtained from the LLs energy in (C). The bandwidth $W = E_{up} + E_{down}$ is also plotted for the experimentally available region. Using $E = (N + 1/2)\hbar\omega_c$ and an effective mass in the order of $\sim 0.4 m_e$ [37], we can estimate the energy $E_{up}$ from the top of the band to vHS$_{up}$ via the hole-like LLs and the energy $E_{down}$ from the bottom of the band to vHS$_{down}$ via the electron-like LLs. For the experimentally accessible region, a bandwidth $W = E_{up} + E_{down}$ as a function of displacement field is extracted and shown. (**F**) Landé g-factor estimation from vHS splitting. Calculated vHS splitting in filling ($\Delta\nu$) as function of displacement field is shown in the colored lines for a range of g*. Experimental $\Delta\nu$ is shown in black.

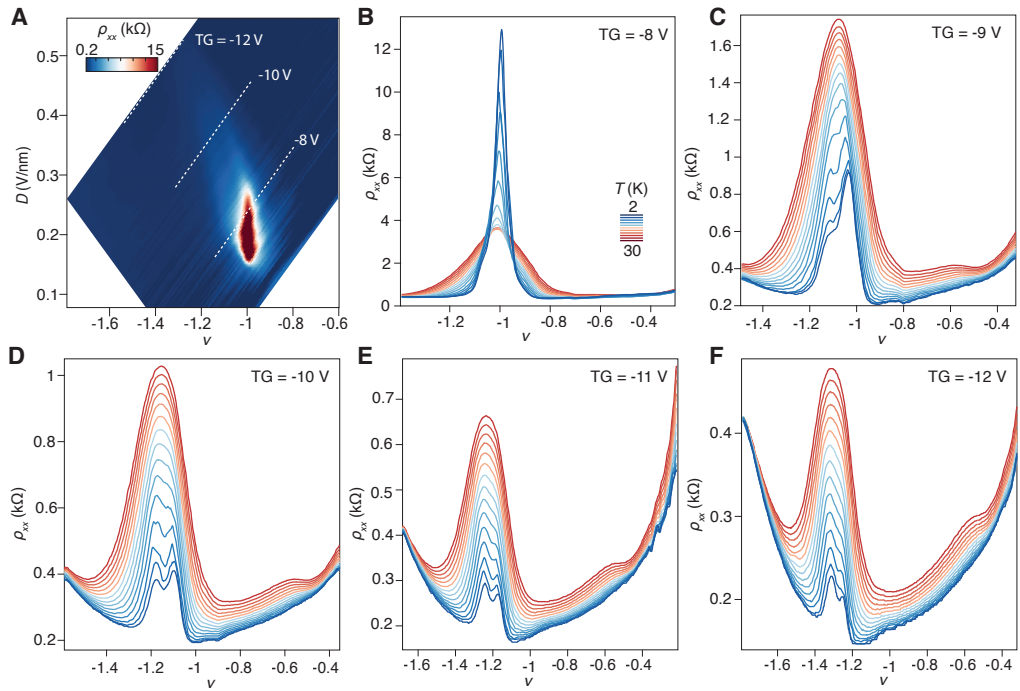

**Fig. S2. Metallic versus insulating behavior**
**(A)** Displacement field versus doping map of resistivity at 1.5 K and 1 T. Lines of constant top gates are also shown. **(B-F)** Temperature dependence at 0 T for different top gates. For TG = -8 V (B), insulating behavior is seen at ν = −1